\documentclass[preprint,10pt]{sigplanconf}

\pdfoutput=1

\usepackage{amsmath}
\usepackage{graphicx}
\usepackage{subcaption}
\usepackage{algorithm}
\usepackage{algpseudocode}
\usepackage[squaren,binary,thinspace,thinqspace,textstyle]{SIunits}
\usepackage{booktabs}

\begin{document}

\graphicspath{{figs/}}

\setlength{\pdfpageheight}{\paperheight}
\setlength{\pdfpagewidth}{\paperwidth}

\conferenceinfo{CONF 'yy}{Month d--d, 20yy, City, ST, Country} 
\copyrightyear{20yy} 
\copyrightdata{978-1-nnnn-nnnn-n/yy/mm} 
\doi{nnnnnnn.nnnnnnn}




\title{A lightweight optimization selection method for Sparse Matrix-Vector
  Multiplication}

\authorinfo{Athena Elafrou\and Georgios Goumas\and Nectarios Koziris}
           {National Technical University of Athens}
           {\{athena,goumas,nkoziris\}@cslab.ece.ntua.gr}

\maketitle

\begin{abstract}
In this paper, we propose an optimization selection methodology for the
ubiquitous sparse matrix-vector multiplication (SpMV) kernel. We propose two
models that attempt to identify the major performance bottleneck of the kernel
for every instance of the problem and then select an appropriate optimization to
tackle it. Our first model requires online profiling of the input matrix in
order to detect its most prevailing performance issue, while our second model
only uses comprehensive structural features of the sparse matrix. Our method
delivers high performance stability for SpMV across different platforms and
sparse matrices, due to its application and architecture awareness. Our
experimental results demonstrate that a) our approach is able to distinguish and
appropriately optimize special matrices in multicore platforms that fall out of
the standard class of memory bandwidth bound matrices, and b) lead to a
significant performance gain of 29\% in a manycore platform compared to an
architecture-centric optimization, as a result of the successful selection of
the appropriate optimization for the great majority of the matrices. With a
runtime overhead equivalent to a couple dozen SpMV operations, our approach is
practical for use in iterative numerical solvers of real-life applications.
\end{abstract}

\category{J-2}{Computer Applications}{Physical Sciences and Engineering}
\category{G-1-3}{Numerical Linear Algebra}{}

\terms
Algorithms, Performance


\section{Introduction}

The ubiquitous sparse matrix-vector multiplication (SpMV) kernel is a
fundamental building block of popular iterative methods for the solution of
sparse linear systems ($Ax=b$), and the approximation of eigenvalues and
eigenvectors of sparse matrices ($Ax=\lambda x$). Such problems arise in a
diverse set of application domains, including large-scale simulations of
physical processes using multi-physics and multi-disciplinary approaches,
information retrieval, medical imaging, economic modeling and others. Optimizing
SpMV has always been a challenging task due to a number of inherent performance
limitations, as a result of the algorithmic nature of the kernel, the employed
sparse matrix storage format and the sparsity pattern of the matrix. SpMV is
characterized by a very low flop:byte ratio, since the kernel performs $O(n^2)$
operations on $O(n^2)$ amount of data, indirect memory references as a result of
storing the matrix in a compressed format, irregular memory accesses to the
source vector due to sparsity, and loop overheads in the presence of a large
amount of very short rows in the matrix~\cite{goumas2009performance}.

Most optimization efforts proposed in the literature over the past few
decades~\cite{cuthill1969reducing, agarwal1992high, temam1992characterizing,
  toledo1997improving, pinar1999improving, im2001optimizing, im2004sparsity,
  mellor2004optimizing, vuduc2005fast, willcock2006accelerating} have focused
either on a subset of the aforementioned performance issues or a single hardware
platform and are, consequently, neither portable nor provide stable performance
gains even within a single hardware platform. This is one of the reasons why
such optimizations are difficult to adopt in real-life applications. Another
reason concerns the non-negligible runtime overhead that usually accompanies
such optimizations. This overhead may include format conversion, format
parameter tuning, reordering the matrix, etc. If the numerical solver does not
have enough iterations to amortize this cost, the benefit of applying the
optimization may be outweighed. To make matters worse, one might incur the
overhead without any guarantee of a performance improvement. Based on all the
previous observations, we have come to believe that the next solid step for
improving SpMV performance no longer involves proposing new and expensive
optimizations, but applying the plethora of available optimizations whenever
they can be effective, i.e., transforming the challenge of \textit{developing
  new optimizations} to \textit{selecting an appropriate optimization for the
  target matrix and architecture}. We claim that a ``universal'' optimization
effort for SpMV is feasible, and, towards this direction, we propose a
methodology to automatically select an efficient optimization for SpMV, that is
both application and architecture aware. Our approach seeks to incorporate the
following key characteristics:
\vspace{-\topsep}
\vspace{0.5em}
\begin{itemize}
\item{\emph{stable}: Optimization selection should provide performance
  improvements for all sparse matrices.}
\item{\emph{cross-platform}: Optimization selection should be beneficial on
  every architecture.}
\item{\emph{lightweight}: The runtime overhead of optimization selection should
  be low in order for it to be applicable in real-life scenarios.}
\end{itemize}
\vspace{-\topsep}
\vspace{0.5em}

In order to provide performance stability, our methodology attempts to detect
the major performance bottleneck of SpMV for the input matrix. We formulate the
decision making as a classification problem in
Section~\ref{sec:methodology:classes}, assuming classes represent performance
bottlenecks. We then develop a profiling-based classifier in
Section~\ref{sec:methodology:classifiers:profiling}, that relies on
micro-benchmarks to classify the matrix. Architectural characteristics are
implicitly deduced through these micro-benchmarks, making this classifier
architecture independent. As the online-profiling phase has a non-negligible
cost, which may outweigh the optimization benefit, we go one step beyond, and
propose in Section~\ref{sec:methodology:classifiers:feature} a classifier that
relies only on structural features of the input matrix, avoiding any online,
profiling-based information. This feature-based classifier is pre-trained during
an offline stage with the use of machine learning techniques, and only performs
feature extraction on-the-fly. The runtime overhead of this classifier is very
low, equivalent to only a couple of dozen SpMV operations, making it extremely
lightweight. Once the prevailing performance bottleneck of a matrix has been
detected, we accordingly apply an optimization that could tackle it. Although we
employ a simple and easy-to-implement set of optimizations, our approach is
compatible to any optimization able to tackle the specific performance
bottlenecks. We have tested our method on one multi-core (Intel Sandy Bridge)
and one many-core (Intel Xeon Phi) architecture for a wide set of candidate
matrices. Our experimental results demonstrate that a) on the multicore
platform, our approach is able to distinguish and appropriately optimize special
matrices that fall out of the standard class of memory bandwidth bound matrices,
and b) on the manycore platform, our approach leads to a significant performance
gain of 29\% compared to an architecture-centric optimization, achieved by the
successful selection of the appropriate optimization for the great majority of
the matrices.


\section{Background}
\label{sec:background}

\begin{figure}[t!]
\begin{center}
  \centering
  \includegraphics[scale=0.74,keepaspectratio]{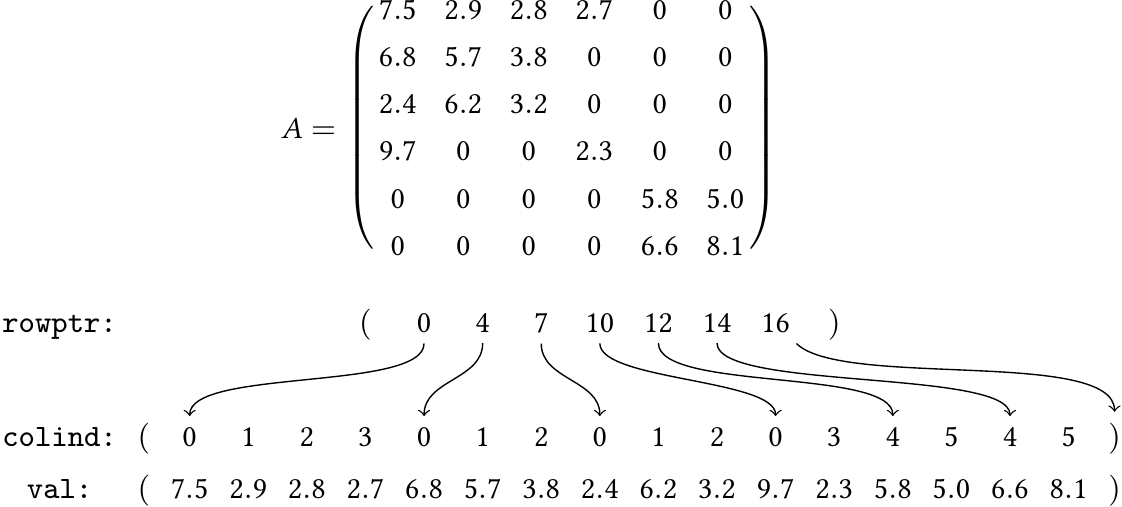}
  \caption{The Compressed Sparse Row (CSR) sparse matrix storage format.}
  \label{fig:csr}
\end{center}
\end{figure}

In order to avoid the extra computation and storage overheads imposed by the
large majority of zero elements contained in a sparse matrix, it has been the
norm to store the nonzero elements of the matrix contiguously in memory and
employ auxiliary data structures for the proper traversal of the matrix and
vector elements. The most widely used format, namely the \textit{Compressed
  Sparse Row} (CSR) format~\cite{saad1992numerical}, uses a row pointer
structure to index the start of each row within the array of nonzero elements,
and a column index structure to store the column of each nonzero element. An
example of this format is given in Figure~\ref{fig:csr}. The SpMV kernel using
this format is given in Algorithm~\ref{algo:spmv}.

\begin{algorithm}
\begin{algorithmic}[1]
  \Procedure{Spmv}{$A$::in, $x$::in, $y$::out}
  \Statex{$A$: matrix in CSR format}
  \Statex{$x$: input vector}
  \Statex{$y$: output vector}
  \For{$i \gets 1,N$}
  \For{$j \gets rowptr[i],rowptr[i+1]$}
  \State{$y[i] \gets val[j]\cdot x[colind[j]]$}
  \EndFor
  \EndFor
  \EndProcedure
\end{algorithmic}
\caption{SpMV computation using the CSR format.}
\label{algo:spmv}
\end{algorithm}

Examining Algorithm~\ref{algo:spmv}, we notice three potential performance
issues for SpMV:

\begin{enumerate}
\item \emph{indirect memory references}: This is the most apparent implication
  of sparsity. Since we only want to store the nonzero elements of the matrix,
  we need auxiliary indexing structures to access them from memory. For the CSR
  format we use the \texttt{colind} and \texttt{rowptr} data
  structures. Indexing, however, introduces additional load operations, traffic
  for the memory subsystem, and cache interference~\cite{pinar1999improving}.
\item \emph{irregular memory accesses to vector $x$}: Access to vector $x$ in
  sparse matrices is irregular and depends on the sparsity structure of the
  matrix. This fact complicates the process of exploiting any spatial or
  temporal reuse in the access to vector
  $x$~\cite{geus1999towards,pichel2004improving}.
\item \emph{short row lengths}: Many sparse matrices contain a large number of
  rows with short lengths. This fact may degrade performance due to the overhead
  incurred by the small trip count of the inner
  loop~\cite{mellor2004optimizing}.
\end{enumerate}

\section{Optimization Selection Methodology}
\label{sec:methodology}

Depending on the sparsity pattern of the matrix and the underlying architecture,
a suitable optimization for SpMV may vary due to varying performance
issues. Thus, SpMV could benefit from an optimization selection process. The
benefits of such an approach are two-fold: firstly, performance can be optimized
for all problem instances, and, secondly, we can leverage previous research on
SpMV that has generally focused on different instances of the problem. The
solution to the optimization selection problem will enable the development of
frameworks that will intelligently select the optimal optimization for a
particular input matrix, thus providing performance stability for the SpMV
kernel.

\subsection{Formulation as a Classification Problem}
\label{sec:methodology:classes}
The optimization selection problem can be solved in various ways. One could
simply take an empirical approach: measure how different optimizations work for
a particular matrix on the target machine and then apply the most efficient
optimization on future runs. However, this would require a heavy profiling phase
with a non-negligible runtime cost, that could outweigh any benefit from
optimizing SpMV. We select a more elegant and lightweight approach to solve the
optimization selection problem. We formulate it as a classification problem, by
assuming that every matrix belongs to a single class, representing its major
performance bottleneck. For every class, we assign a corresponding optimization
that attempts to tackle the specific bottleneck. Given an input sparse
matrix, its class is predicted and the corresponding optimization is applied. In
this context, we define the following classes:

\begin{itemize}
\item{\emph{CML}: This class refers to matrices that suffer from excessive LLC
  misses and can therefore be bound by cache miss latencies. The source of these
  misses can be determined by examining the SpMV kernel in
  Algorithm~\ref{algo:spmv}. Accesses to the matrix elements (\texttt{val}) and
  indexing data structures (\texttt{rowptr, colind}) show a regular streaming
  behavior, which can be easily detected and prefetched by hardware and, thus,
  are of limited concern to us. However, accesses to the input vector are
  performed through indirect indexing, which requires special hardware
  prefetching mechanisms not available in current architectures. Thus, if the
  sparsity pattern of the matrix is very irregular, latency can become a
  bottleneck for SpMV.}
\item{\emph{MB}: This class includes matrices that have saturated the available
  memory bandwidth. This is the dominating class for SpMV on modern multicore
  architectures due to its very low flop:byte ratio.}
\item{\emph{IMB}: This class appears mostly on many-core architectures, where
  the large number of threads exposes highly uneven row lengths in the matrix or
  regions with completely different sparsity patterns, resulting either in
  workload imbalance or computational unevenness.}
\item{\emph{CMP}: This class includes matrices that are bound by
  computation. Such matrices are mostly matrices that fit in the system's caches
  and are, therefore, not limited by main memory bandwidth.}
\end{itemize}

The above classes are quite generic, as they serve our initial goal, which is to
categorize a matrix based on its prevailing performance bottleneck. Each class
covers a wide variety of matrices with different sparsity patterns. Thus, one
could further sub-categorize each class, based on more distinguishing structural
features. This would serve a more elaborate optimization selection scheme and is
left for future work.

Instead of defining our own classes we could have used cluster analysis to
discover groups of matrices that are similar in some manner. However, it seemed
more intuitive to leverage our own experience and the extensive research that
has been realized over the past few decades on SpMV optimization, and define our
own classes. A ``blind'' machine learning approach is not necessary in this
case, since the major performance bottlenecks of the kernel have already been
identified, granting us a basic understanding of the problem. Another option
would be to directly define optimizations as classes. Every class would simply
represent the best optimization for its group of matrices. However, in this
scenario, we would have to associate features of a matrix to every optimization
in our search space instead of every performance bottleneck, which could be more
error-prone. Also, this approach cannot lead to a cross-platform decision
making, as it is optimization dependent and optimizations can be completely
different from one hardware platform to another.

\subsection{Classifiers}
\label{sec:methodology:classifiers}
We follow two approaches to solve our classification problem and guide the
optimization selection for SpMV. The first approach involves a hand-tuned
classification algorithm that relies on performance data collected during an
online profiling phase of the input matrix. We will henceforth refer to this
classifier as the \textit{profiling-based} classifier. The second approach
leverages a machine learning technique to train a classifier on a predefined
set of matrices during an offline stage and only requires a small number of
structural features to be extracted from the input matrix on-the-fly. This is
the \textit{feature-based} classifier.

\subsubsection{Profiling-Based Classifier}
\label{sec:methodology:classifiers:profiling}
In this approach, the class of a matrix is determined through a series of
micro-benchmarks that are executed on the input matrix during an online
profiling phase. These benchmarks attempt to implicitly extract the impact of
the architectural characteristics of the underlying hardware platform on SpMV
performance. We define three benchmarks that run a CSR-based SpMV kernel with
some modification:

\begin{itemize}
\item{\emph{noxmiss}: This benchmark tries to eliminate irregular memory
  accesses to vector x, by setting the column indices of all nonzero elements to
  zero through the \texttt{colind} array of CSR. Since irregularity results in
  cache misses, it is indicative of matrices that belong to the CML class.}

\item{\emph{inflate}: This benchmark uses 64-bit indices for the
  \texttt{rowptr} and \texttt{colind} arrays of CSR in order to increase the
  working set size of SpMV (the size of the indexing structures is actually
  doubled). We expect this benchmark to indicate which matrices are hindered by
  an increase in the working set size, and, consequently, have a high
  probability of being memory bandwidth bound.}

\item{\emph{balance}: This benchmark measures the execution time of all
  individual threads and reports the arithmetic mean. This metric is indicative
  of the performance that could be reclaimed by balancing the load. We must note
  here that our default partitioning scheme for SpMV is static one-dimensional
  row partitioning, with an equal (or as close to equal as possible) number of
  nonzero elements per partition.}
\end{itemize}

Each benchmark reports SpMV execution time. For a given input matrix, we run all
three benchmarks along with the baseline CSR-based implementation and compute
the speedup of each benchmark over the baseline CSR, with the exception of
\textit{inflate}, for which we compute the inverse. The classification is
performed by evaluating the impact of each benchmark on the baseline
implementation. First, we select the benchmark with the highest speedup and, if
it exceeds some threshold, we classify the matrix accordingly. If not, we
examine the benchmark with the second highest speedup and so on and so forth. If
none of the three benchmarks has a noteworthy impact on performance, then the
matrix is labeled as CMP. The thresholds have been empirically set for the time
being taking into account the ability of a micro-benchmark to provide a naive or
realistic bound on the speedup that can be achieved for SpMV by tackling the
corresponding problem. For example, the \textit{noxmiss} benchmark, which
completely eliminates irregularity in the matrix, provides a naive performance
bound and, thus, requires a strict (high) threshold.

\subsubsection{Feature-Based Classifier}
\label{sec:methodology:classifiers:feature}
We also use a machine learning approach to build a classifier. The advantage of
machine learning in this case is that, given a set of classes, the
classification rules can be automatically deduced based on a training data set,
and, subsequently, be used for predictions, requiring no profiling executions.

We experiment with a Decision Tree classifier and a Naive Bayes classifier and
employ supervised learning algorithms to generate them. The classifiers are
built based on features extracted from the matrix structure. Decision Tree
learning is performed using an optimized version of the CART algorithm, which
has a runtime cost of $O(n_{features}$\\$n_{samples}\log n_{samples})$ for the
construction of the tree, where $n_{features}$ is the number of features used
and $n_{samples}$ is the number of samples used to build the classifier. The
Naive Bayes classifier is created using Maximum A Posteriori (MAP) training with
a Gaussian distribution assumption, which has a time complexity of
$O(n_{features}n_{samples})$. The query time for these classifiers is $O(\log
n_{samples})$ and $O(n_{classes}n_{features})$ respectively. We generate both
classifiers using the scikit-learn machine learning toolkit.

\subsubsection*{Feature Extraction}
This classifier uses real-valued features to perform the
classification. Table~\ref{table:features} shows all the features we
experimented with, along with the time complexity of their extraction from the
input matrix. Most features are parameters that represent comprehensive
characteristics of a sparse matrix and are closely related to SpMV
performance. For example, the number of nonzero elements per row can indicate
load imbalance during SpMV execution. If a matrix contains very uneven row
lengths and a row-partitioning scheme is being employed for workload
distribution, then threads that are assigned long rows will take longer to
execute resulting in thread imbalance. Uneven row lengths can easily be detected
through the corresponding metrics in Table~\ref{table:features} (see
$nnz_{\{min,max,avg,sd\}}$). Similarly, irregularity in the accesses to the
input vector, which can lead to excessive cache misses, can be estimated by
examining how spread out the elements in every row are (see
$dispersion_{\{avg,sd\}}$).

\begin{table}
  \scriptsize
  \begin{center}
    \begin{tabular}{|c|c|c|}
      \hline
      \textbf{Feature} & \textbf{Definition} & \textbf{Complexity} \\
      \hline
      $size$ & 0:exceeds or 1:fits in LLC & $\Theta(1)$ \\
      $density$ & $\frac{NNZ}{N*M}$ & $\Theta(1)$ \\
      $nnz_{\text{min}}$ & $\min\{nnz_{1},\dotsc,nnz_{N}\}$ & $\Theta(N)$ \\
      $nnz_{\text{max}}$ & $\max\{nnz_{1},\dotsc,nnz_{N}\}$ & $\Theta(N)$ \\
      $nnz_{\text{avg}}$ & $\frac{1}{N}\sum_{i=1}^{N}nnz_{i}$ & $\Theta(N)$ \\
      $nnz_{\text{sd}}$ & $\sqrt{\frac{1}{N}\sum_{i=1}^{N} (nnz_{i}-nnz_{avg})^2}$ & $\Theta(2N)$ \\
      $bw_{\text{min}}$ & $\min\{bw_{1},\dotsc,bw_{N}\}$ & $\Theta(N)$ \\
      $bw_{\text{max}}$ & $\max\{bw_{1},\dotsc,bw_{N}\}$ & $\Theta(N)$ \\
      $bw_{\text{avg}}$ & $\frac{1}{N}\sum_{i=1}^{N}bw_{i}$ & $\Theta(N)$ \\
      $bw_{\text{sd}}$ & $\sqrt{\frac{1}{N}\sum_{i=1}^{N} (bw_{i}-bw_{avg})^2}$ & $\Theta(2N)$ \\
      $dispersion_{\text{avg}}$ & $\frac{1}{N}\sum_{i=1}^{N}disp_{i}$ & $\Theta(N)$ \\
      $dispersion_{\text{sd}}$ & $\sqrt{\frac{1}{N}\sum_{i=1}^{N} (disp_{i}-dispersion_{avg})^2}$ & $\Theta(2N)$ \\
      $clustering$ & $\frac{1}{N}\sum_{i=1}^{N}clust_{i}$ & $\Theta(NNZ)$ \\
      $miss\_ratio$ & $\frac{1}{N}\sum_{i=1}^{N}misses_{i}$ & $\Theta(NNZ)$ \\
      \hline
    \end{tabular}
  \end{center}
  \caption{Sparse matrix features used for classification. $N$ denotes
    the number of rows in the matrix, $M$ the number of columns and
    $NNZ$ the number of nonzero elements. $nnz_{i}$ is the number of
    nonzero elements of row $i$, $bw_{i}$ the column distance between
    the first and last nonzero element of row $i$, $disp_{i} =
    \frac{nnz_{i}}{distance_{i}}$, $clust_{i} =
    \frac{ngroups_i}{nnz_{i}}$, where $ngroups_i$ is the number of
    groups formed by consecutive elements in row $i$ and, finally,
    $misses_i$ is the number of nonzero elements in row $i$ that can
    generate cache misses. We naively say that an element will
    generate a cache miss when its distance from the previous element
    in the same row exceeds the cache line size of the system.}
  \label{table:features}
\end{table}

\subsection*{Training Data Selection}
We train our classifiers with a data set consisting of sparse matrices from the
University of Florida Sparse Matrix Collection~\cite{davis2011university}. We
have selected a matrix suite consisting of 115 matrices from a wide variety of
application domains, to avoid being biased towards a specific sparsity
pattern. We must note here that we are examining all matrix sizes, not only
matrices exceeding the system's LLC size. Also, we have decided not to balance
the training set in terms of class representation, as that would require
redefining the training set on every hardware platform (as all classes are not
equally represented on each platform), which is not desirable for a
cross-platform approach. We did, nevertheless, experiment with balanced training
sets, but the improvement in the accuracy of our classifiers was not significant
enough to justify imposing such a limitation.

\subsection*{Training Data Labeling}
Labeling refers to the process of assigning a class to each matrix that will be
used in the training process of our classifier. Since the class of a matrix
cannot be determined in a straightforward manner, we use our profiling-based
classifier for this purpose. An issue that arises by this choice, and cannot be
ignored, is the validity of the labelings, as it affects the accuracy of the
trained classifier. Since the decision rules in the profiling-based classifier
have been empirically set, a matrix can be mislabeled if it falls on the
boundaries that separate classes.

\subsection{Optimization Pool}
A multitude of optimization techniques and specialized formats have been
proposed in the literature for improving the performance of SpMV. Most of them
require a preprocessing phase, either to modify the sparsity pattern of the
matrix itself or to generate new data structures for the representation of the
matrix, that result in higher compression ratios of its memory footprint, better
load balancing of the workload among threads, etc. This preprocessing overhead
needs to remain low, as it may offset the benefit of applying the optimization,
in particular when only tens of iterations are required for a numerical solver
to converge.

As most legacy code in scientific applications uses the standard CSR format for
sparse computations, we decided, for the sake of simplicity and in order to
minimize the preprocessing costs, to incorporate only CSR-based
optimizations. We currently experiment with a single optimization per class,
given in Table~\ref{table:optimizations}. However, our methodology is not tied
to specific optimizations and can be easily extended to incorporate any
optimization proposed in the literature, as long as it is guaranteed to be
effective for the corresponding class.

For the CML class of matrices, we employ one of the most important techniques
for tolerating increasing cache miss latencies, e.g., prefetching. Since the
source of excessive cache misses in SpMV is a result of indirect memory
addressing on the input vector, which cannot be efficiently tackled by current
hardware prefetching mechanisms, software prefetching was applied. A single
prefetch instruction was inserted in the inner loop of SpMV, with a fixed
prefetch distance (although it can be tuned on a matrix basis for a higher
performance gain). The prefetch distance was set equal to the number of elements
that fit in a single cache line of the hardware platform, assuming
double-precision floating-point values. The intuition behind this choice is that
for very irregular matrices, there will be limited to no spatial locality for
accesses to $x$ in a single row, thus requiring a prefetch instruction per
element in order to hide the latency. The data is prefetched in the L1
cache. For matrices that are memory-bandwidth bound, we employ a rather simple
compression scheme, just for illustrative purposes. We use delta indexing on the
column indices of the nonzero elements of the matrix, a technique which was
originally applied on SpMV by Pooch and Nieder~\cite{pooch1973survey}. We use 8-
or 16-bit deltas wherever possible, but never both, in order to limit the
branching overhead during SpMV computation. For matrices that suffer from load
imbalance we employ either the \textit{dynamic} or \textit{auto} scheduling
policies available in the OpenMP runtime system~\cite{dagum1998openmp}. When the
\textit{auto} schedule is specified, the decision regarding scheduling is
delegated to the compiler, which has the freedom to choose any possible mapping
of iterations (rows in this case) to threads.  Finally, for compute-bound
matrices we combine inner loop unrolling with vectorization.

\begin{table}
  \small
  \begin{center}
    \begin{tabular}{|c|c|}
      \hline
      \textbf{Class} & \textbf{Optimization} \\
      \hline
      CML & software prefetching on vector x \\
      MB & column index compression through delta coding~\cite{pooch1973survey}\\
      IMB & \textit{auto} or \textit{dynamic} scheduling (OpenMP) \\
      CMP & inner loop unrolling + vectorization \\
      \hline
    \end{tabular}
  \end{center}
  \caption{Mapping of matrix classes to targeted optimizations.}
  \label{table:optimizations}
\end{table}

\section{Experimental Evaluation}
\label{sec:experiment}
Our experimental evaluation focuses on two aspects: matrix classification
accuracy and the overall benefit of applying our methodology for optimizing
SpMV.

\subsection{Experimental Setup and Methodology}
Our experiments are performed on two hardware platforms, including an Intel Xeon
Phi coprocessor (codename Xeon Phi) and a cache-coherent non-uniform memory
access (cc-NUMA) multiprocessor system. The NUMA system is a four-way eight-core
Intel Xeon E5-4620 configuration (codename Sandy Bridge-EP). In the following,
we will refer to each platform using its codename. Table~\ref{table:hw} lists
the technical specifications of our test platforms in more detail.

Both systems were running a 64-bit version of the Linux OS. For the compilation
of the software involved in our performance evaluation, we used ICC-15.0.0,
while we used the OpenMP parallel programming API. To simulate the typical
sparse matrix storage case, we used 64-bit, double precision floating point
values for the nonzero elements.

\begin{table}
  \scriptsize
  \centering
  \begin{tabular}{cccc}
  \toprule
  \textbf{Codename} & \textbf{Xeon Phi} & \textbf{Sandy Bridge-EP} \\
  \midrule
  \textbf{Model} & Intel Xeon Phi 3120P & Intel Xeon E5-4620 \\
  \textbf{Microarchitecture} & Intel Many Integrated Core & Intel Sandy Bridge \\
  \textbf{Clock frequency} & 1.10 GHz & 2.26 GHz \\
  \textbf{L1 cache (D/I)} & \unit{32}{\kibi\byte}/\unit{32}{\kibi\byte} &
  \unit{32}{\kibi\byte}/\unit{32}{\kibi\byte} \\
  \textbf{L2 cache} & \unit{512}{\kibi\byte} \textit{(per core)} & \unit{256}{\kibi\byte} \textit{(per core)}\\
  \textbf{L3 cache} & - & \unit{16}{\mebi\byte} \\
  \textbf{Cores/Threads} & 57/228 & 8/16 \\
  \midrule
  \multicolumn{3}{c}{\textbf{Multiprocessor configurations}} \\
  \midrule
  \textbf{Processors} & 1 & 4 \\
  \textbf{Cores/Threads} & 57/228 & 32/64 \\
  \textbf{Sustained memory b/w} & \unit{119}{\gibi\byte\per\second} &
  4$\times$\unit{13.5}{\gibi\byte\per\second} \\
  \bottomrule
  \end{tabular}
  \caption[Technical characteristics of the hardware platforms used for the
    experimental evaluations.]{Technical characteristics of the hardware
    platforms used for the experimental evaluations. The sustained memory
    bandwidth figures are obtained with the STREAM
    benchmark~\cite{mccalpin1995stream} utilizing the full system.}
  \label{table:hw}
\end{table}

\subsection{Classifier Accuracy}
\label{sec:accuracy}
First, we evaluate our feature-based classifier in terms of accuracy, assuming
the labels generated by the profiling-based classifier are correct. The
profiling-based classifier is not eligible to such an analysis, as we currently
have no way of accurately deciding upon the major performance bottleneck of a
matrix and, thus, correctly evaluating this classifier. Using optimizations to
determine the label of a matrix would require more than one optimization per
class in order to be safe, so we do not currently follow this approach. However,
we do provide later on some insight into how the classifiers perform compared to
the best among the employed optimizations for every matrix.

We estimate how accurately our models will perform in practice using
Leave-One-Out cross validation. According to this methodology, for a training
set of $k$ matrices (115 in our case), $k$ experiments are performed. For each
experiment $k-1$ matrices are used for training and one for testing. The final
accuracy reported by the whole process is the fraction of $k$ matrices for which
the classifier correctly predicts their class, i.e. the prediction matches the
original labeling of the matrix. This metric is important, as it determines the
suitability of the selected optimization for the target matrix. However, it is
not decisive, contrary to standard classification problems, in the sense that it
depends on the correct labeling of each matrix, which is not straightforward in
some cases. There exist matrices that have competing performance bottlenecks,
and, thus, their labeling is ambiguous. This fact, actually, allows us to
tolerate a less accurate labeling algorithm, such as our profiling-based
classifier. It also means that our classifier might improve the performance of
SpMV even in the case of a ``misprediction'', as long as this misprediction
corresponds to another performance bottleneck.  Thus, achieving the highest
possible accuracy is desirable, but not of the highest priority.

Tables~\ref{table:classifiers:phi} and~\ref{table:classifiers:sand} give an
overview of the final feature-based classifiers for each hardware platform under
consideration along with the time complexity of the feature extraction and the
achieved accuracy. The selection of features for these classifiers has been a
result of exhaustive search. Decision Tree classifiers seem to work slightly
better for our problem, with a maximum accuracy of 77\% on Xeon Phi and 84\% on
Sandy Bridge-EP. However, the Naive Bayes classifiers manage to attain
competitive accuracies using fewer features.

\begin{table}
  \scriptsize
  \begin{center}
    \vspace{-\topsep}
    \begin{tabular}{|c|p{2.5cm}|c|c|}
      \hline
      \textbf{Classifier} & \textbf{Feature sets} & \textbf{Complexity} & \textbf{Accuracy(\%)}\\
      \hline
      Decision Tree & $size, bw_{\{avg,sd\}}$ $nnz_{\{min,max,avg,sd\}}$ $miss\_ratio$ $dispersion_{sd}$ & O(NNZ) & 77\\
      \hline
      Naive Bayes & $nnz_{\{min,max,sd\}}$ $bw_{avg}$ $dispersion_{\{avg,sd\}}$ & O(N) & 75\\
      \hline
    \end{tabular}
  \end{center}
  \caption{Feature-based classifiers for Xeon Phi.}
  \label{table:classifiers:phi}
\end{table}

\begin{table}
  \scriptsize
  \begin{center}
    \vspace{-\topsep}
    \begin{tabular}{|c|p{2.5cm}|c|c|}
      \hline
      \textbf{Classifier} & \textbf{Feature sets} & \textbf{Complexity} & \textbf{Accuracy(\%)}\\
      \hline
      Decision Tree & $size, bw_{\{avg,sd\}}$ $nnz_{\{min,max,avg,sd\}}$ $dispersion_{sd}$ $miss\_ratio$ & O(NNZ) & 84\\
      \hline
      Naive Bayes & $size, nnz_{\{min,max\}}$ & O(N) & 83\\
      \hline
    \end{tabular}
  \end{center}
  \caption{Feature-based classifiers for Sandy Bridge-EP.}
  \label{table:classifiers:sand}
\end{table}

\subsection{Optimization Selection Performance}
Contrary to standard classification problems, our key measure of success is not
classifier accuracy---rather, we aim to maximize the average performance
improvement of optimization selection over pre-selecting the most
straightforward optimization on each architecture. In this way, we designate the
value of optimization selection for SpMV. For our multi-core architecture (Sandy
Bridge-EP) we consider column index compression (see
Table~\ref{table:optimizations}) to be the straightforward optimization technique,
while on our many-core architecture (Xeon Phi), which is characterized by very
wide SIMD units, we consider vectorization to be the straightforward
optimization.

\begin{figure*}[]
  \begin{subfigure}[!]{\textwidth}
    \centering
    \includegraphics[scale=0.35,keepaspectratio]{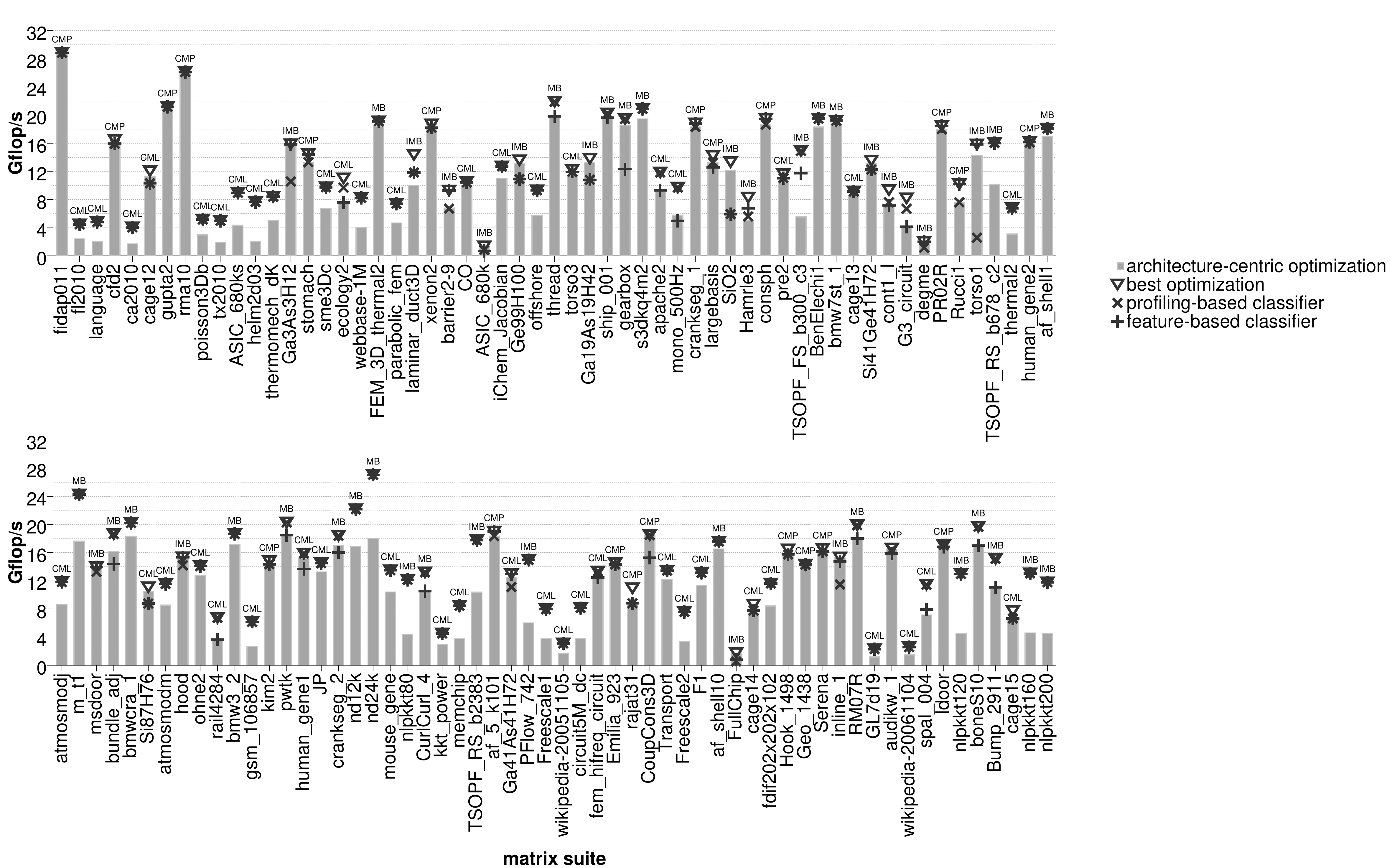}\hspace{-11em}
    \raisebox{9mm}{\includegraphics[scale=0.35,keepaspectratio]{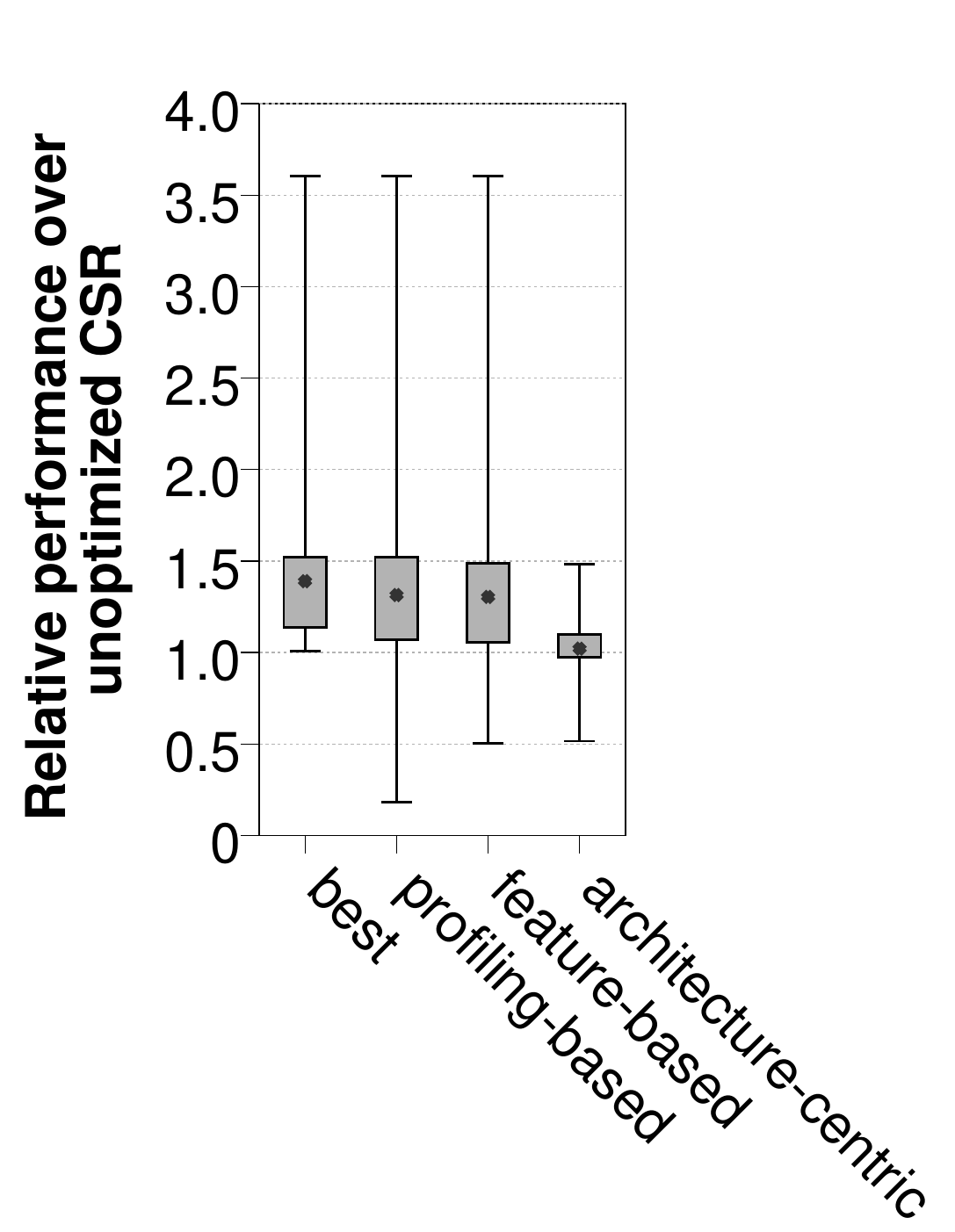}}
    \label{fig:phi}
    \caption{Xeon Phi (224 threads)}
  \end{subfigure}
  \begin{subfigure}[!]{\textwidth}
    \centering
    \includegraphics[scale=0.35,keepaspectratio]{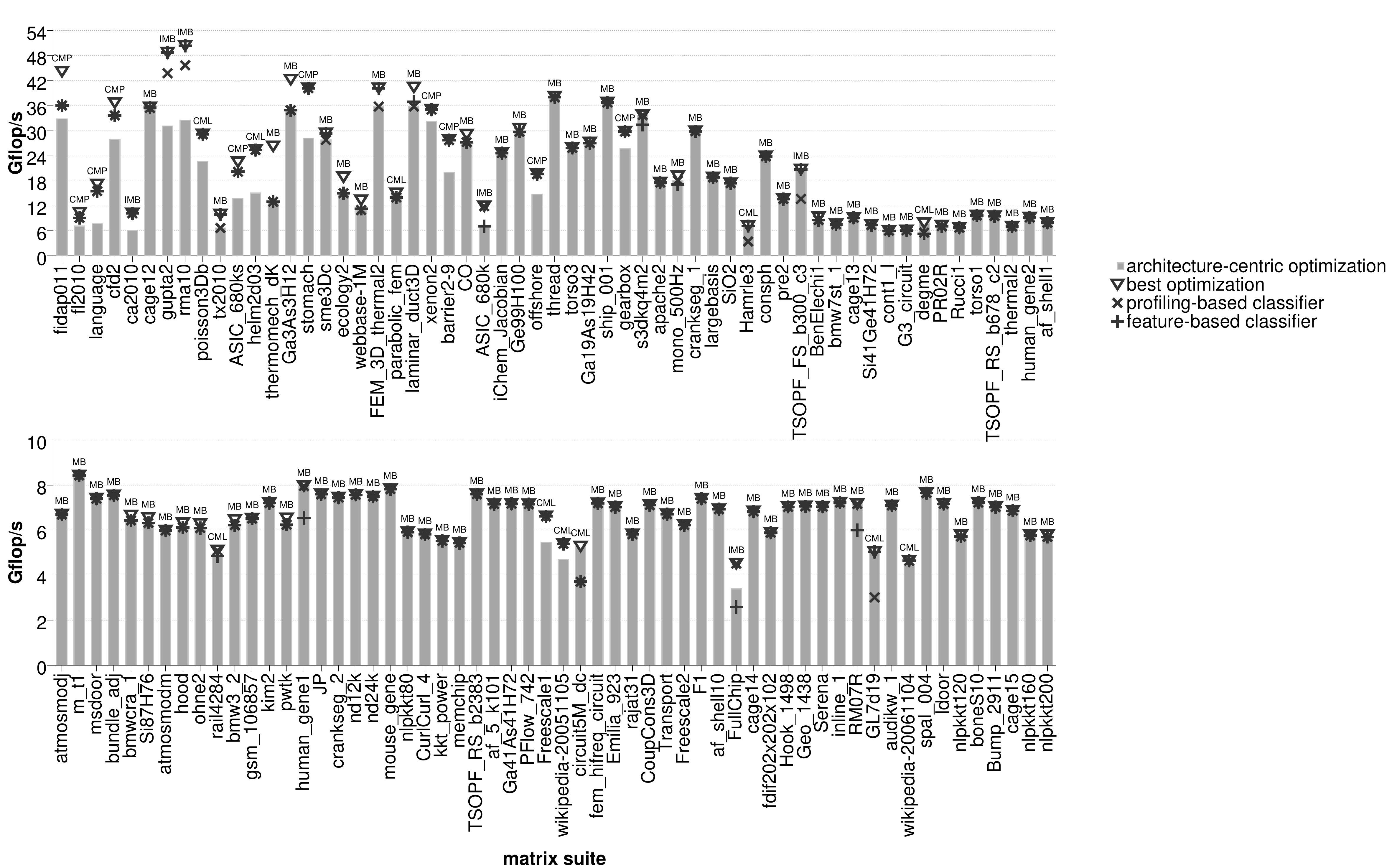}\hspace{-11em}
    \raisebox{9mm}{\includegraphics[scale=0.35,keepaspectratio]{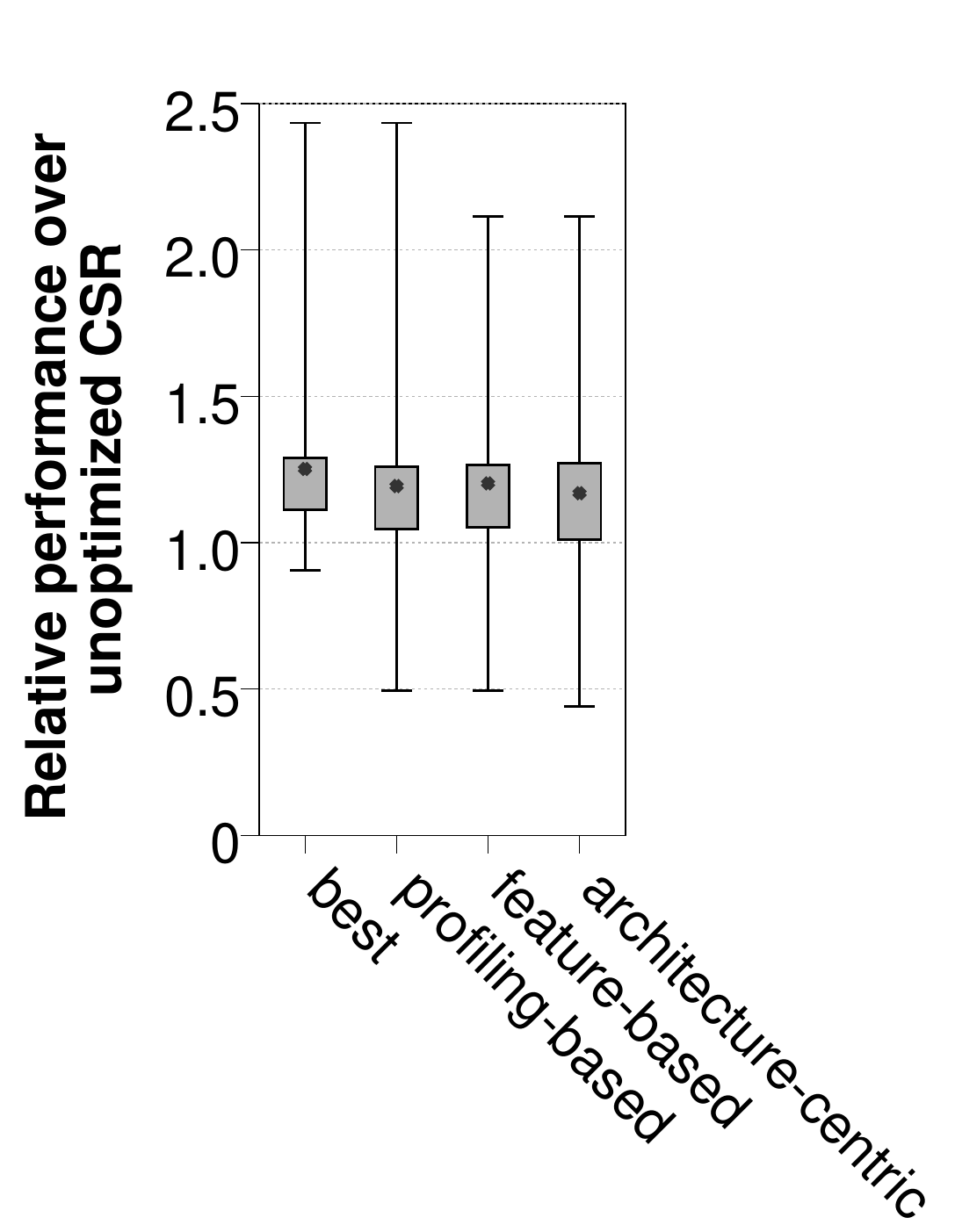}}
    \caption{Sandy Bridge-EP (32 threads)}
    \label{fig:sandy}
  \end{subfigure}
  \caption{Performance results of optimization selection on Xeon Phi and Sandy
    Bridge-EP. \textit{architecture-centric optimization} refers to the
    optimization corresponding to the CMP class for Xeon Phi and MB class on
    Sandy Bridge-EP, while \textit{best optimization} represents the maximum
    performance achieved with any of the tested optimizations (see
    Table~\ref{table:optimizations}).}
  \label{fig:performance}
\end{figure*}

Figure~\ref{fig:performance} presents the raw SpMV performance achieved by the
architecture-centric optimization, as defined previously, the best optimization
among our pool of optimizations, and the optimization selected by each of our
classifiers. The labeling of each matrix according to the profiling-based
classifier is also provided. We note that predictions from the feature-based
classifier are obtained once again through Leave-One-Out cross-validation, as
described in the previous subsection. On each platform we use the feature-based
classifier with the highest accuracy reported in
Tables~\ref{table:classifiers:phi} and~\ref{table:classifiers:sand}. Our initial
observation concerns the diversity of performance bottlenecks on our
experimental platforms, demonstrated by the labeling of the matrices. On Xeon
Phi, all classes are equally represented, unlike Sandy Bridge-EP, which is
dominated by matrices belonging to the MB class, as expected. Our classifiers
manage to successfully capture this trend due to their architecture awareness.

Taking a closer look at Figure~\ref{fig:performance}, we notice that the
profiling-based classifier does not always match the best optimization,
indicating that it might be misclassifying matrices. However, this is not always
the case, as some matrices may be correctly classified, but the corresponding
optimization is not as effective. This is the case with \textit{Ga3As3H12} and
\textit{torso1} on Xeon Phi. On this platform, mislabelings happen mostly with
matrices that are competing between belonging to the CML or IMB
class. Nonetheless, in some cases, e.g., \textit{Hamrle3, Rucci1, torso1,
  Ga41As41H72, rajat31, inline\_1}, our feature-based classifier manages to
select a better optimization, even though it has been trained with labels
generated by the profiling-based classifier. On Sandy Bridge-EP, where the
majority of the matrices belong to the MB class, both of our classifiers manage
to optimize most of the corner cases, e.g., \textit{cfd2, gupta2, rma10,
  offshore}, etc. On this platform, most
matrices not belonging to the MB class fit in the aggregate cache of the system,
thus exposing weaknesses in the computational part of SpMV.

To get an overall evaluation of our approach, for each matrix we
define the following metrics:
\begin{itemize}
\item{\textit{ideal classifier speedup}: the speedup provided by applying the
  best optimization amongst those under consideration}
\item{\textit{profiling-based classifier speedup}: the speedup provided by
  applying the optimization selected by the profiling-based classifier}
\item{\textit{feature-based classifier speedup}: the speedup provided by
  applying the optimization selected by the feature-based classifier}
\end{itemize}

Using the unoptimized CSR-based SpMV implementation as the baseline, for each of
the aforementioned metrics we plot the average, min and max values over the
entire matrix set, as well as the interquartile range. i.e., the range between
the first and third quartiles, which represents the 25\%-75\% range of the
values. Figure~\ref{fig:performance} presents the corresponding results on each
platform in the form of a box plot. The lower and upper whiskers correspond to
the min and max values, while the bottom and top of the box are the first and
third quartiles respectively. The circle corresponds to the average.

On Xeon Phi, the importance of optimization selection is more prominent,
providing an average speedup of $1.31\times$ with the profiling-based classifier
and $1.30\times$ with the feature-based classifier, reaching $94.2\%$ and
$93.5\%$ respectively of the best possible gain with the applied
optimizations. This is mainly due to the architectural characteristics of the
Xeon Phi coprocessor, which result in a quite balanced representation of each
class. The coprocessor has a large amount of cores, which favor workload
imbalance, and a very expensive (an order of magnitude bigger compared to
multi-cores) cache miss latency, which further exposes irregularity in the
accesses to the input vector. Thus, there are many matrices that exceed the LLC
size and are not memory bandwidth bound. On the other hand, Sandy Bridge-EP is
dominated by memory bandwidth bound matrices and can, therefore, only benefit
from the correct classification of matrices belonging to minority classes. This
makes higher accuracy rates more important on this architecture. Overall, our
classifiers manage to optimize most of those matrices, leading to a $1.19\times$
and $1.20\times$ average speedup respectively and adding a 3\% and 4\%
improvement over the architecture-centric optimization. Even though there are
some misclassifications, they can be partially tolerated by the fact that the
architecture-centric optimization can also be ``harmful'' for some matrices.

In total, the potential of an optimization selection approach for SpMV can be
better estimated if we assume we have a ``best'' classifier, i.e., a classifier
that always selects the best optimization presented in
Figure~\ref{fig:performance}, and compare it to an ``architecture-centric''
optimizer. In the box plots provided, we see that the ``best'' classifier
improves SpMV performance over the ``architecture-centric'' by 37\% on Xeon Phi
and 9\% on Sandy Bridge-EP on average.

It is important to point out that the achieved performance improvements
presented in this work are limited by the optimizations we have selected. As
this is not an actual optimization framework, we selected optimizations that can
be easily implemented. Higher performance gains can be attained by a more
elaborate examination of the wide variety of optimizations found in the
literature and their association to our matrix classes. Selecting an even more
suitable optimization after the matrix has been classified, based on its
distinguishing structural features, is left for future work.

\subsection{Runtime overhead}
Our profiling-based optimization selection approach comprises two steps: running
a number of micro-benchmarks on the input matrix and then applying our
empirically-tuned classification algorithm to select the appropriate
optimization. On the other hand, the online stage of the feature-based
optimization selection methodology comes down to extracting the required
features from the input matrix and then predicting its class using the
pre-trained classifier.

\begin{table}
  \small
  \begin{center}
    \begin{tabular}{|c|c|c|}
      \hline
      & \textbf{profiling-based} & \textbf{feature-based} \\
      \hline
      Xeon Phi & 462 & 14 \\
      Sandy Bridge-EP & 462 & 16\\
      \hline
    \end{tabular}
  \end{center}
  \caption{Average runtime overhead of optimization selection expressed in
    multi-threaded CSR-based SpMVs, using both classifiers presented in
    Section~\ref{sec:methodology:classifiers}.}
  \label{table:overhead}
\end{table}

The runtime overhead of both classification approaches was measured in
multithreaded CSR SpMV operations and is presented in
Table~\ref{table:overhead}, i.e., we report the ratio
$\frac{T_{classification}}{T_{SpMV}}$, where $T_{classification}$ is the
execution time of the classification and $T_{SpMV}$ the execution time of a
single multithreaded SpMV operation. Apparently, the feature-based classifier
has a significant advantage over its profiling-based alternative, as it requires
only a couple dozen operations, which can be easily amortized in the context of
an iterative numerical solver. We must note here that this overhead does not
include any preprocessing required for applying an optimization. Among the
optimizations applied in our evaluation, only the one for MB matrices requires
preprocessing.

\section{Related Work}
\label{sec:related}

Different sparse matrices have different sparsity patterns, and different
architectures have different strengths and weaknesses. In order to achieve the
best SpMV performance for the target sparse matrix on the target platform, an
autotuning approach has long been considered to be beneficial. The first
autotuning approaches attempted to tune parameters of specific sparse matrix
storage formats. Towards this direction, the Optimized Sparse Kernel Interface
(OSKI) library~\cite{vuduc2005oski} was developed as a collection of high
performance sparse matrix operation primitives on single core processors. It
relies on the SPARSITY framework~\cite{im2004sparsity} to tune the SpMV kernel,
by applying multiple optimizations, including register blocking and cache
blocking. Autotuning has also been used to find the best block and slice sizes
of the input sparse matrix on modern CMPs and GPUs~\cite{choi2010model}.

There have been some research efforts closer to our work. The clSpMV
framework~\cite{su2012clspmv} is the first framework that analyzes the input
sparse matrix at runtime, and recommends the best representation of the given
sparse matrix, but it is restricted to GPU platforms. Towards the same
direction, the authors in~\cite{guo2014performance} present an analytical and
profile-based performance modeling to predict the execution time of SpMV on GPUs
using different sparse matrix storage formats, in order the select the most
efficient format for the target matrix. For each format under consideration,
they establish a relationship between the number of nonzero elements per row in
the matrix and the execution time of SpMV using that format, thus encapsulating
to some degree the structure of the matrix in their methodology. Similarly,
in~\cite{li2015performance}, the authors propose a probabilistic model to
estimate the execution time of SpMV on GPUs for different sparse matrix
formats. They define a probability mass function to analyze the sparsity pattern
of the target matrix and use it to estimate the compression efficiency of every
format they examine. Combined with the hardware parameters of the GPU, they
predict the performance of SpMV for every format. Since compression efficiency
is the determinant factor in this approach, it is mainly targeted for memory
bandwidth bound matrices. Closer to our approach is the SMAT autotuning
framework~\cite{li2013smat}. This framework selects the most efficient format
for the target matrix using feature parameters of the sparse matrix. It treats
the format selection process as a classification problem, with each format under
consideration representing a class, and leverages a data mining approach to
generate a decision tree to perform the classification, based on the extracted
feature parameters of the matrix. The distinguishing advantage of our
optimization selection methodology over the aforementioned approaches, is that
it decouples the decision making from specific optimizations, by predicting the
major performance bottleneck of SpMV instead of SpMV execution time using a
specific optimization. Thus, in contrary to the above frameworks, where
incorporating a new optimization requires either retraining a model or
defining a new one, our decision-making approach allows an autotuning framework
to be easily extended, simply by assigning the new optimization to one of the
classes.

\section{Conclusions \-- Future Work}
\label{sec:conclusion}

In this paper we presented a cross-platform methodology for optimizing the SpMV
kernel and establish that, depending on the sparsity pattern of the matrix and
the underlying architecture, a suitable optimization can be selected to improve
SpMV performance. We formulate optimization selection for SpMV as a
classification problem, with each class corresponding to a performance
bottleneck. We first propose a profiling-based classifier, that relies on
benchmarking the input matrix to perform the decision making. We also leverage
machine learning to train Decision Tree and Naive Bayes classifiers that use
comprehensive structural features to perform the classification and rely on the
profiling-based classifier for the training process.  Experimental evaluation on
115 sparse matrices and two platforms has demonstrated that our methodology is
very promising, especially for the latest many-core architectures, which
intensify various performance issues of the SpMV kernel.

Concerning future work, another direction that needs to be explored is whether a
multi-level optimization approach would be beneficial for SpMV. It is very
likely that, once the major performance issue of a matrix has been successfully
addressed, another bottleneck will be exposed. Actually, the features that are
extracted from the matrix and fed to our featured-based classifier can also be
used in a following step to select even more targeted optimizations, depending
solely on the sparsity pattern. We also intend to experiment with other machine
learning techniques, in order to further improve the accuracy of our
feature-based classifier. For example, semi-supervised learning, which allows
some of the training data to be unlabeled, could be helpful, since there are
matrices whose class cannot be easily determined. Finally, we plan to test our
approach on GPU platforms.






\bibliographystyle{abbrvnat}
\bibliography{adapt16}





\end{document}